\documentclass[preprint, tighten, twocolumn]{aastex62}
\usepackage{apjfonts}
 
\newcommand{\etal}{et al.}
\newcommand{\hal}{H$\alpha$}
\newcommand{\kms}{km s$^{-1}$}

\begin{document} 

\title{No Evidence for Periodic Variability in the Light
  Curve of Active Galaxy J0045+41}

\shorttitle{No Periodicity in J0045+41}
\shortauthors{Barth \& Stern}

\author[0000-0002-3026-0562]{Aaron J. Barth}
\affiliation{Department of Physics and Astronomy, 4129 Frederick
  Reines Hall, University of California, Irvine, CA, 92697-4575, USA}

\author[0000-0003-2686-9241]{Daniel Stern}
\affiliation{Jet Propulsion
  Laboratory, California Institute of Technology, 4800 Oak Grove
  Drive, Mail Stop 169-221, Pasadena, CA 91109, USA}

\begin{abstract}

Dorn-Wallenstein, Levesque, \& Ruan recently presented the
identification of a $z=0.215$ active galaxy located behind M31 and
claimed the detection of multiple periodic variations in the object's
light curve with as many as nine different periods. They interpreted
these results as evidence for the presence of a binary supermassive
black hole with an orbital separation of just a few hundred AU, and
estimated the gravitational-wave signal implied by such a system. We
demonstrate that the claimed periodicities are based on a
misinterpretation of the null hypothesis test simulations and an error
in the method used to calculate the false alarm probabilities. There
is no evidence for periodicity in the data.
  
\end{abstract}


\section{Introduction}

It is generally expected that binary supermassive black holes should
form in the aftermath of galaxy mergers, but direct evidence for close
(sub-pc separation) binary black holes in galaxy nuclei remains
elusive despite much observational effort.  One possible way to
identify binary black holes is to search for quasars or active
galactic nuclei (AGN) exhibiting periodic variations in brightness
that might be associated with the binary orbital period. Periodic flux
variability may be expected as a result of the interaction between the
binary and a circumbinary accretion disk
\citep[e.g.,][]{macfadyen2008, haiman2009}, or through periodic
modulation caused by relativistic boosting of emission from gas bound
to the lower-mass secondary black hole in an unequal-mass binary
\citep{dorazio2015}.

Over the past few years there have been numerous claims for periodic
flux variations in AGN identified through searches of large samples of
AGN light curves obtained in time-domain surveys
\citep{graham2015a,graham2015b,liu2015,liu2016,charisi2016}.  Genuine
periodic flux variability in AGN is apparently extremely rare, and it
is debatable whether any objects have yet shown compelling evidence of
light curve periodicity, but the number of candidates is growing
rapidly.  False positive detections of periodicity are a serious risk,
because random, aperiodic AGN variability can sometimes produce
repeated up-and-down variations that can easily be mistaken for
periodicity, particularly when only a few cycles are seen in the data
\citep{vaughan2016}. However, once candidates are identified, further
long-duration monitoring can determine whether the periodicity is likely
to be real or not. To test the significance of any possible
periodicity identified in an AGN light curve, it is essential to carry
out careful Monte Carlo simulations to estimate the false positive
rate: that is, the probability that a suitably chosen stochastic
variability process will produce a periodicity signal at least as
strong as that found in the data.

Recently, \citet[][hereinafter DLR]{dw17} presented the identification
of a previously unknown AGN, J0045+41. This object appears
  within the disk of M31 as projected on the sky, and has been
  classified in the past as an M31 globular cluster or globular
  cluster candidate
  \citep[e.g.,][]{galleti2004}.\footnote{This object appears
    in the SIMBAD catalog as 2MASS J00452729+4132544.  In NED, it is
    listed under the designation CXO J004527.3+413255. \citet{kim2007}
    identified it as a ``possible'' globular cluster in M31, but noted
    that some of the possible clusters in their sample might actually
    be background galaxies.  Assuming the globular cluster
    identification for this object, \citet{maccarone2016} interpreted
    the associated X-ray source as an X-ray binary, based on
    observations from \emph{NuSTAR} and \emph{Swift}.} DLR examined
the X-ray spectrum of J0045+41 and obtained an optical spectrum
showing that the object is actually a broad-lined AGN at $z=0.215$.
They also reported the detection of multiple periodicities in the
object's optical light curve with the strongest periodicity at 354.84
days, and interpreted the results as evidence for a binary
supermassive black hole with an orbital separation of just a few
hundred AU.

Given the rarity of periodicity in AGN and the importance of
identifying binary black holes, the claim of compelling evidence for
multiple periodicities in J0045+41 by DLR should be examined
carefully. If correct, it would be a remarkable discovery, and if this
object were even merely a promising candidate for periodic variability
it would merit extensive follow-up observations to verify or rule out
the periodicity. The purpose of this paper is to show that the claims
of periodicity in J0045+41 are unfortunately not valid. The
periodicities are spurious, arising from a combination of incorrect
interpretation of null hypothesis simulation tests and an incorrect
calculation of the false alarm probability.

\section{Evaluating the evidence for periodicity}

DLR use $g$-band light curve data from the Palomar Transient Factory
(PTF) to search for periodicity in this object.  The light curve is
shown in their Figure 7. It is immediately apparent that the light
curve is very noisy and sparsely sampled with large seasonal gaps. The
data points and error bars in the plot are so crowded together that it
is difficult to see most of the individual points.  DLR do not list
the light curve data in a table, nor do they give any quantitative
information about the uncertainties on the data points, and at present
only a portion of this light curve is available in the PTF public
archive. Nevertheless, it is clear from inspection that both the AGN
variability amplitude and the S/N of the light curve are rather
low. There is a gradual increase in the source brightness over the
6-year monitoring duration, with the last three monitoring seasons
having a slightly higher mean brightness (by perhaps $\sim0.25$ mag)
than the earlier seasons.  From the outset, it is unlikely that data
of this quality would be sufficient to demonstrate any periodicity to
high significance, let alone multiple periodicities. PTF $r$-band data
are also presented, with higher S/N than the $g$-band data, but they
primarily consist of just two monitoring seasons and the $r$-band
light curve does not appear to show any obvious variability.

To measure the periodogram of the $g$-band light curve, DLR use an
algorithm that employs a non-parametric periodic model rather than the
common approach of using a sinusoidal model \citep[e.g.,][]{graham2015b,
  charisi2016, liu2016}. The periodogram is sampled on a uniform grid
of 2000 trial periods ranging from 60 to 1000 days. As shown in their
Figure 8, they find a trend of rising power $P$ at longer timescales,
but with irregular peaks and troughs in the plot of $P$ versus period
$T$. Such features can occur due to the sampling cadence of the light
curve rather than the variability power of the AGN itself.  The
important question is whether any of the peaks represent power
significantly above what might be expected due to the ordinary
stochastic variability of an AGN. This requires Monte Carlo
simulations to test the null hypothesis, which in this case is the
hypothesis that the observed periodogram power can be generated by an
aperiodic variability process.

To perform this test, DLR carry out simulations in which they generate
mock light curves following a damped random walk (DRW) process with
DRW parameters selected to match those of the observed light curve,
then resample the light curves and add noise to match the cadence and
S/N of the data, and then measure the periodogram of each simulated
light curve. In itself, this is a standard and reasonable approach to
carrying out null hypothesis simulations (with a few caveats mentioned
at the end of this section).  The results of the DRW simulations are
illustrated in the left panel of their Figure 10, which compares the
periodogram of the data with that of the simulations, plotted as the
mean DRW periodogram and a $\pm1\sigma$ confidence band around the
mean based on the distribution of power at each period in the
simulated periodograms.

This plot shows that the periodicity power in the data is
  within or below the $1\sigma$ confidence band of the DRW simulations
  at nearly all the trial periods ranging from 60 to 1000 days. There
  is only one narrow peak where the data power exceeds the $1\sigma$
  confidence band of the simulations, corresponding to the claimed
  $\sim355$ day period.\footnote{The amount by which the 355
      day peak exceeds the $1\sigma$ confidence band of the
      simulations is not listed in the text of the paper and cannot be
      determined accurately from the figure, but it appears to be only
      slightly larger than a $1\sigma$ deviation from the mean DRW
      power.}  However, a feature extending to slightly above
  $1\sigma$ in a small range of adjacent period bins (out of 2000
  periods searched) could presumably arise very easily by chance in an
  aperiodic DRW light curve.  When searching for a possible
  periodicity signal over a broad range of periods without \emph{a
    priori} knowledge of where the periodicity might lie, the
  look-elsewhere effect comes into play
  \citep[e.g.,][]{gross2010}, and a peak would have to be a
  very strong outlier extending far above the $1\sigma$ confidence
  band of the simulations in order to have a chance of being truly
  significant. For the J0045+41 light curve, since the level of power
  over the entire period search range can evidently be reproduced by
  typical simulations of stochastically variable light curves, the
  most straightforward interpretation of the null hypothesis test
  simulations is that there is simply no evidence of periodicity.

Despite the fact that there is only one peak exceeding the mean
  power level of the DRW simulations by more than $1\sigma$ (and only
  by a small amount), DLR claim to detect evidence of \emph{nine}
  separate periodicity peaks in the data. To test quantitatively
  whether a claimed peak in the periodogram is significant or not, the
  simulation results can be used to calculate the expected false
  positive rate or false alarm probability (FAP). The FAP essentially
gives a $p$-value for the null hypothesis, that is, the probability
that the aperiodic DRW simulations can produce a periodogram peak as
great or greater than one seen in the data.  DLR calculate the FAP in
the following way. The total number of DRW simulations performed is
$N_\mathrm{DRW} = 96,000$.  They first rebin the periodogram into
$N_\mathrm{trial} = 100$ period bins containing 20 periods per
bin. Within each bin, they find the highest value of the periodogram
power in the data signal [power $P_S(T)$ at period $T$], and then
determine the number of simulated DRW periodograms that exceed
$P_S(T)$ within that period bin; this quantity is called
$N_\mathrm{DRW}(>P_S(T))$.  The ratio $N_\mathrm{DRW}(>P_S(T)) /
N_\mathrm{DRW}$ then gives the fraction of all simulations that exceed
the data power within that period bin, i.e., a local $p$-value
for that particular bin. To account for the look-elsewhere
  effect, DLR divide this ratio by $N_\mathrm{trial}$, calculating the
  FAP as
\begin{equation}
  \mathrm{FAP} = \frac{N_\mathrm{DRW}(>P_S(T))}{N_\mathrm{DRW}\times
    N_\mathrm{trial}}.
  \label{eqn1}
\end{equation}

However, this is not a correct method to calculate the FAP.
The nature of the look-elsewhere effect is that the true FAP must be
\emph{greater} than the $p$-value calculated by considering just the
single bin where the peak is found in the data, given the presence of
many other possible period bins where a DRW simulation could randomly
generate a periodogram peak.  Dividing by $N_\mathrm{trial}$ in
Equation \ref{eqn1} makes the derived FAP 100 times \emph{smaller}
than the local $p$-value calculated from the single bin in
which a peak is found, and it will always yield
  $\mathrm{FAP}\leq0.01$ in every period bin, even in the extreme case
  of a bin in which \emph{every} simulation exceeds the periodogram
  power of the data. The premise of dividing the single-bin $p$-value
by $N_\mathrm{trial}$ to account for the look-elsewhere effect is
fundamentally incorrect, and this error leads to an enormous
overestimate of the significance of the putative periodicities.

How should the FAP be determined for a candidate peak in the data?
Since there is no \emph{a priori} information on where a peak might be
found in the periodogram, calculating the FAP essentially amounts to
determining the fraction of the randomly generated simulations having
power at \emph{any} period that would appear to be at least as
significant as the observed peak. In other words, the FAP is a global
$p$-value for the putative peak.  If the noise level of the simulated
periodograms were uniform across the full range of trial periods, then
this would simply require finding the fraction of all simulations that
reach power levels at least as high as the observed power of the
candidate peak. However, the noise level in the simulated periodograms
is strongly period-dependent. In this situation, the global $p$-value
cannot be determined by using a uniform power threshold across all
possible periods. Instead, it is necessary to examine the distribution
of simulated power at each trial period. Let $p_\mathrm{peak}$ be the
local $p$-value of the candidate peak, equal to the fraction of
simulations that exceed the data power in the period bin where the
peak is found. Then, in each other period bin, a fraction
$p_\mathrm{peak}$ of the simulations will locally appear to have power
at least as significant as that of the candidate peak. Let
$N_\mathrm{sig}$ be the number of simulations that reach this
significance threshold in at least one period bin, and
$N_\mathrm{tot}$ be the total number of simulations performed. The FAP
or global $p$-value is then $N_\mathrm{sig}/N_\mathrm{tot}$.  This
method ensures that the FAP will be greater than or equal to the
$p$-value calculated from the single bin where the peak is found, as
required. As an additional check, the FAP can be recomputed using a
range of possible bin widths for the periodogram, to test whether the
derived FAP is sensitive to the choice of period bin size.

The results of the incorrect FAP calculation are listed in Table 1 and
shown in the the right panel of Figure 10 of DLR. In this figure, they
plot $(1 - \mathrm{FAP})$ as a function of period, i.e., the
  likelihood that the observed power indicates a real detection of
  periodicity in the source.  The plotted values of $(1 -
  \mathrm{FAP})$ exceed 0.99 at \emph{all} periods, since the
  calculated FAP must be $<0.01$ in every period bin according to
  Equation \ref{eqn1}.  In other words, this plot implies a $>99\%$
confidence detection of periodicity in \emph{every} period bin within
the search range. This is not a physically plausible result and it is
certainly incompatible with the low quality of the light curve data;
it is a clear warning sign that the method used to calculate the FAP
is not valid. For a light curve of this quality, any reasonable
calculation of FAP should result in values of $(1-\mathrm{FAP})$ just
above zero at all or nearly all periods.

DLR identify nine separate ``peaks'' in the plot of $(1 -
\mathrm{FAP})$ corresponding to the claimed periodicities, but only
one of these peaks (at $T=354.84$ days) corresponds to a location
where the periodogram exceeds the $1\sigma$ band of the simulations,
and at least one of the peaks has power substantially below the mean
level of the DRW simulations.\footnote{Table 1 of DLR lists the
  derived periods for these peaks to five significant figures (e.g.,
  354.84 days, or an implied precision of $\sim14$ minutes), but they
  do not give uncertainties on the periods.}  Table 1 lists FAP values
for these peaks (to six significant figures) ranging from
$1.01854\times10^{-3}$ to $7.78917\times10^{-3}$. Although these might
appear to be moderately significant results, Figure 10 shows that
\emph{every} period bin ostensibly has FAP $<10^{-2}$ (as a
  result of the incorrect Equation \ref{eqn1}), so there is little
difference between the FAP at the putative peaks and the FAP at any
other point in the periodogram. This again is a warning sign that the
FAP has not been calculated correctly.

From the data presented in Table 1 of DLR and Equation 1, it is
evident that for the strongest claimed peak at $T\approx355$ days, the
single-bin $p$-value is 0.10, meaning that 10\% of the DRW simulations
exceeded the periodicity power in the data within the period bin where
the peak was found. The look-elsewhere effect guarantees that the
actual FAP must be substantially greater than 0.10, so the null
hypothesis is a very plausible explanation for the observed power in
this peak. This conclusion is in stark contrast to the results of DLR,
who found FAP $=1.0\times10^{-3}$ for this peak using the incorrect
Equation 1. For the other peaks, the single-bin $p$-values range from
0.33 to 0.78, strongly contradicting the notion that these might be
significant peaks. While we cannot determine the actual false positive
rates for these peaks without the full DRW simulation results or
carrying out new null hypothesis simulations, the simulation results
presented by DLR are already sufficient to demonstrate that there is
no good evidence for periodicity at all. The FAP values listed in
Table 1 of DLR are too low by at least a factor of 100 in every case.

In Figure 12, DLR show phase-folded light curves that illustrate six
of the identified periods. Examination of these plots gives a clearer
understanding of how these spurious periodicities arise as a result of
the sparse temporal sampling of the data and an aliasing effect
exacerbated by the large amount of freedom in the shape of the
phase-folded light curves. The ``supersmoother'' periodogram algorithm
used by DLR allows for somewhat arbitrarily shaped bumps and kinks in
the phase-folded light curves, in contrast to the smooth and regular
variability of a sinusoid.  Given this freedom to generate light
curves of almost any shape, the period-finding algorithm is finding
apparent periodicities for which the lower-flux data points from early
in the campaign and the higher-flux data points from late in the
campaign are almost entirely segregated in phase from each other.  For
example, in the phase-folded light curve for the $T\approx355$ day
period (which is claimed to be the strongest periodicity), phases
0.0--0.4 are dominated by photometry from the first half of the
campaign while phases 0.5--0.9 are dominated by photometry from the
second half of the campaign.  This makes the phase-folded light curve
appear to show that phases 0.0--0.4 have lower flux than phases
0.5--0.9, when the actual underlying behavior is that the AGN had
lower flux during the first three years and higher flux during the
final three years of monitoring.  Furthermore, the strongest
individual feature in the phase-folded light curve is a cluster of
points between phases 0.8 and 0.9 (shown in light green) that were all
obtained in a single year's observations. This feature represents a
single, brief brightening event in the AGN light curve with no
evidence that this feature was periodic or recurring in any other
year.  This cannot in itself give useful evidence for periodicity. A
convincing period detection would require data points from different
observing seasons in the light curve to be more thoroughly mixed in
phase, such that a variability feature could be clearly seen to repeat
over multiple cycles. The same considerations apply to all of the
phase-folded light curves for the various putative periods, which show
similar phase segregations between the early (low-flux) and late
(high-flux) points in the light curve. All of these false
periodicities would be easily excluded by a correctly implemented FAP
calculation.

Two additional points made by \citet{vaughan2016} bear repeating here.
First, there is evidence that the DRW model is not a perfect
description of AGN variability, particularly on short timescales. For
example, high-quality \emph{Kepler} light curves \citep{mushotzky2011}
demonstrate power spectrum slopes of $\alpha\sim-2.6$ to $-3.3$ (for
$P\propto f^\alpha$ where $f$ is the temporal frequency of
variations), steeper than the $\alpha=-2$ high-frequency behavior of a
DRW, and this difference could lead to an overestimate of the
significance of a periodicity signal if the data were compared only
with DRW simulations. Ideally, in a case where evidence for
periodicity appears strong, simulations should be carried out against
a realistic range of power spectral shapes using the method of
  \citet{timmer1995}, or other more flexible variability models such
  as the CARMA model \citep{kelly2014}.  Second, an accurate
assessment of the light curve uncertainties is very important. If the
magnitude uncertainties on light curve data points are overestimated,
this will lead to injection of too much white-noise variability into
the simulated light curves when simulated data points are adjusted to
account for measurement errors. This would also tend to give an
overestimate of the periodicity significance.

DLR's search for periodicity in J0045+41 was motivated by the
suggestion of a 76-day periodicity in $B$ and $V$-band light curves
reported by \citet{vilardell2006} as part of a search for eclipsing
binaries in M31. The \citet{vilardell2006} data are very sparsely
sampled, having been obtained in one observing run per year over a
five-year duration, with each observing run spanning 3--5 consecutive
nights. Such sampling is not at all sufficient to detect a 76-day
period.  Their phase-folded light curve (also reproduced in Figure 13
of DLR) clearly demonstrates strong variability over the five
monitoring seasons with an amplitude of $\sim1$ mag in the $B$
band. In the phase-folded light curve, the data points are grouped
into five widely separated groupings in phase with large gaps between
them. The explanation for the derived periodicity at $T\approx76$ days
is simply that these five groupings correspond to the five observing
runs spread over five years. The AGN varied significantly from year to
year, and the period-finding algorithm just rearranged the five yearly
groups of data points in a way that resembled one period of a
quasi-sinusoidal cycle.  Almost any AGN light curve, even one that was
monotonically increasing in flux over 5 years, could be rearranged to
appear as though it were periodic in a phase-folded light curve if it
were sampled with this extremely sparse cadence. Vilardell
\etal\ point out this aliasing issue in their paper, noting that
periodicities greater than five days in their sample are unlikely to be
real.  There is no reason to interpret the \citet{vilardell2006} data
as providing any evidence of periodicity at all, particularly after
the object was correctly identified as an AGN by DLR.\footnote{The
  \citet{vilardell2006} phased light curve is available in the CDS
  VizieR online catalog associated with their paper, and object
  J0045+41 is object 3571 in their catalog of variable stars (their
  Table 4). The object is not mentioned in the text of their paper,
  presumably because they did not consider it a significant detection
  of periodicity.  They note that ``it is clear that many of the
  period determinations are just an alias introduced by the window
  function, especially for those over five days.'' Among the variable
  objects listed in Table 4 of Vilardell \etal\ having apparent
  periods greater than five days, 39\% have listed periods between 75
  and 90 days, illustrating that the window function of their
  observations leads to numerous spurious periodicities in this range.
  DLR also state that Vilardell \etal\ classified J0045+41 as an
  eclipsing binary, but this is not the case. While Vilardell
  \etal\ found this object as part of a search for eclipsing binaries,
  it is not listed with any classification in their tables. Objects
  identified as eclipsing binaries or Cepheids are listed as such in
  their Table 4, but these amount to only 22\% of their sample of 3964
  variable objects, and the majority of the variable sources were not
  classified. }

DLR also discuss a newly obtained Gemini-N optical spectrum of
J0045+41 as providing possible corroborating evidence for a
supermassive black hole binary.  The broad Balmer lines are asymmetric
with higher flux in the blue wing, and DLR characterize the line
asymmetry in terms of a Gaussian component blueshifted by $\sim4800$
\kms\ with respect to the systemic component. They interpret this
velocity offset as possibly suggestive of a supermassive black hole
binary, an outflow, or a hot spot in the accretion disk of a single
supermassive black hole. For comparison, it has long been known that a
subset of AGN exhibit broad-line asymmetries or velocity offsets
similar to that seen in J0045+41.  These sources are generally
interpreted as being members of the class of double-peaked emitters,
whose line profiles are well modeled as originating from an asymmetric
Keplerian accretion disk around a single black hole
\citep{eracleous1994, strateva2003}.  There is ongoing interest in
using broad-line asymmetries or double-peaked structure as  potential
signatures of supermassive black hole binaries
\citep[e.g.,][]{shen2010, nguyen2016, runnoe2017}, but there is not
yet any direct evidence implicating binarity as the underlying cause
for any observed line-profile features in AGN.  While somewhat
unusual, the \hal\ profile asymmetry in J0045+41 is not particularly
extreme, and it can plausibly be understood within the context of
well-established models for AGN broad-line emission without the need
to invoke a supermassive black hole binary.

\section{Conclusions}
For all of the reasons described above, J0045+41 should not be
considered as a periodically varying AGN, or even as a candidate for a
periodically variable AGN based on the available data. The false alarm
probabilities given by DLR are underestimated by at least a factor of
100 for all of the claimed periodicities. The data and the null
hypothesis simulations do not demonstrate any evidence of periodicity
whatsoever.  In light of these conclusions, there is no reason to
consider this object as a candidate binary black hole or to speculate
about the purported binary's orbital parameters or the anticipated
gravitational-wave strain, as discussed in Section 5 of
DLR.\footnote{We note that the orbital separations listed in Table 2
  of DLR appear to have been calculated without correcting the
  observed-frame periods for cosmological time dilation.}

There has been increasing effort devoted to finding binary black holes
in AGN in recent years, including searches for periodic variations in
photometric light curves and in the radial velocity shifts of broad
emission lines. Finding even one object exhibiting compelling evidence
for a binary supermassive black hole would be an exciting discovery.
The increasing level of attention devoted to binary black holes
following the detection of stellar-mass black hole mergers with LIGO
\citep{ligo2016} will undoubtedly increase the motivation to identify
binary supermassive black holes, and upcoming time-domain surveys
including ZTF and LSST will provide a treasure trove of data in which
to search for periodically varying AGN that may be candidate
binaries. Well-sampled light curves with long durations and high S/N
will be the key to determining whether genuinely periodic AGN can
actually be found.

However, as emphasized by \citet{vaughan2016}, there is a real risk of
finding false periodicities in time-domain data. It is easy for a
stochastic red-noise process to mimic a few cycles of periodic
variability, and searches for periodic AGN must take great care to
avoid false detections. This is true even when well-sampled, high-S/N
data are available.  Any claims of periodicity must be viewed
  with skepticism, and provisionally accepted only if it can be
  demonstrated that the putative periodicity is extremely unlikely to
  have arisen by chance in a stochastic variability process. False
periodicity detections can lead to a variety of negative consequences,
including misdirected investments of telescope time and human effort
in following up spurious results, erroneous inferences for the
expected black hole merger rate and gravitational-wave event rates,
and incorrect results being disseminated to the public through news
releases. The case of J0045+41 provides yet another reminder of the
necessity of carrying out well-designed, reproducible tests that have
the capability to falsify an incorrect hypothesis. Careful attention
to these issues will help future authors (and referees) to be better
equipped to resist the siren call of false periodicities.

\acknowledgements

Research by A.J.B. is supported in part by NSF grant AST-1412693. The
work of DS was carried out at the Jet Propulsion Laboratory,
California Institute of Technology, under a contract with NASA.

\end{document}